\documentclass[ a4paper,twocolumn, 10pt, aip, rsi]{revtex4-1}
\usepackage{amsmath}
\usepackage{graphicx}
\usepackage[outdir=./]{epstopdf}
\usepackage{hyperref}
\usepackage[utf8]{inputenc}

\usepackage[usenames,dvipsnames]{color}
\usepackage[normalem]{ulem}

\begin{document}

\begin{abstract}
We have developed a system for contactless measurement of nonlinear conductivity in the radio-frequency band, and over a wide temperature range. A non-resonant circuit is used to electrically excite the sample, and the induced signal is detected by a resonant circuit whose natural frequency matches higher harmonics of the excitation. A simple modification of the probe allows non-resonant detection suitable for stronger signals. Two measurement procedures are proposed that allow significant excitation power variation, up to 150 W. The apparatus has been validated trough the measurement of the nonlinear response at the superconducting transition of a high-Tc superconductor, and the nematic transition of an iron pnictide.
\end{abstract}

\title{Contactless measurement of nonlinear conductivity in the radio-frequency range\\}

\author{Marija Došlić}
\author{Damjan Pelc}
\author{Miroslav Požek}
\affiliation{Department of Physics, Faculty of Science, University of Zagreb, Bijenička 32, 10000 Zagreb, Croatia}

\maketitle
\section{Introduction}

Conductivity measurements are of great importance in the characterization of materials, especially strongly correlated electron systems, such as high-T$_c$ superconductors. The change of both DC and AC conductivity of such materials is a direct indication of a feature in their phase diagram. In conventional materials the current to voltage ratio is constant, because the field inside a conductor is always sufficiently low to induce a linear response. Nolinear I-V characteristics imply nontrivial processes, and are therefore interesting to study. If the applied voltage is sinusoidal, the resulting current in a system with nonlinear characteristics will deviate from a pure sine wave. It is customary and convenient to express this deviation as a series of higher harmonics, to which underlying physical processes can often be attributed. Although nonlinear response is ubiquitous in condensed matter systems -- from glass-forming liquids to charge- and spin-density wave materials -- it has been somewhat neglected due to experimental difficulties. Our non-contact setup, however, enables simple and reliable nonlinear conductivity measurements for a wide scope of relevant materials, thus being a step towards making the method more commonly used. 

The prevailing method of measuring DC and low-frequency (nonlinear) conductivity is the van der Pauw Method, which uses four contacts, two of which deliver the current, and two that pick-up the voltage drop across the sample. Aside from problems arising from securing contacts on small and sensitive samples, this method has a major drawback at higher frequencies, where the contacts introduce unpredictable impedances. This irreproducible effect can be limited by applying pressure on the contacts, which may result in mechanical damage to the sample.\cite{Miller} At higher frequencies it is therefore better to conduct contactless measurements based on induction and detection of currents in the sample.

The induced signal grows with the frequency of the excitation, therefore a higher frequency is preferable when measuring on small samples, or when detecting small conductivity changes. As the dimension of the conductors leading the signal to and from the sample approaches the wavelength of the signal (for frequencies $\sim$10~MHz), additional reflections and signal distortions arise due to impedance mismatches\cite {Hegman}. At even higher frequencies ($\sim$ GHz), the skin depth of the sample typically diminishes, and the measurements reflect surface, or local, characteristics, instead of bulk properties.\cite{S.C.Lee, Mircea}

Additional challenges appear when measuring nonlinear signals directly. At low frequencies current-induced thermal oscillations occur, and it is difficult to distinguish intrinsic nonlinear effects from signals induced by periodic temperature change. In the microwave range heating is negligible, but the measurement of each harmonic requires a unique probe tuned both to the fundamental frequency and one of its multiples, which renders the experiments inflexible. 

We present an apparatus for direct measurements of nonlinear conductivity in the radio-frequency range. The probe design is fairly straightforward and based on the induction and detection of currents in the sample by a pair of coaxial coils. The electronic circuits can easily be adapted to operate on any specific frequency in the radio-frequency band from $\sim 1$~MHz up to several hundred MHz. We also discuss two different measurement and data analysis methods, one which uses an RF lock-in amplifier, and the other which uses a commercial nuclear magnetic resonance (NMR) spectrometer.

\section{Apparatus and measurement}

The following section is organized into several subsections. First, the induction and detection of currents in the sample are considered for the simplest case of two coaxial coils, without reference to their specific arrangement in our apparatus. The details of probe construction and the electric circuits are given in the second subsection. In subsections \ref{low} and \ref{high} two different measurement procedures are proposed, and their respective weaknesses and strengths discussed. Finally, in section \ref{measurement}, we describe the testing and validating of the apparatus on two samples, where nonlinear response was expected.

\subsection{Operating principle}

Let us consider two coaxial coils wound around a sample. Through one of them -- the excitation coil -- flows a driving current of frequency $\omega$,
\begin{equation}
I=I_0 \sin{\omega t},
\end{equation}
and it creates an in-phase magnetic field in the sample. In our experimental setup the current has an amplitude modulation of frequency $\Omega$, so the actual amplitude is given by 
\begin{equation}
I_0 = I_0'\sin{\Omega t},
\end{equation}
but for reasons of simplicity we will omit this dependence from subsequent calculations. An electrical field is induced in the sample:
\begin{equation}
E\sim \frac{dB}{dt}\sim E_0 \omega \cos{\omega t}.
\end{equation}
As a result, currents are induced. In general their form is nonlinear, and given by:
\begin{equation}
\mathbf{j}=\sigma_1\mathbf{E}+\sigma_3\mathbf{E^3} +...
\end{equation}
where even harmonics don't occur in the vast majority of cases because of time reversal symmetry. The quantities $ \sigma_1$ and $\sigma_3$ are the linear and nonlinear conductivities, respectively. Through trigonometric relations, we obtain: 
\begin{equation}
j=\sigma_1' E_0 \omega \cos{\omega t}+ \sigma_3' {E_0}^3 \omega^3 \cos{3\omega t}+...
\end{equation}
These circular currents cause a magnetic field along the $z$ axis, that in turn induces a voltage in the detection coil. The resulting current through the coil has a $3\omega$ component
\begin{equation}
j\sim {I_0}^3 \omega^4 \sigma_3' \sin{3\omega t}.
\end{equation}
Generally, the conductivity $\sigma_3$ is a complex quantity, meaning that the signal can be both in and out of phase with the excitation. Of course, the above considerations are not limited to the case of conducting samples -- the relevant excitation can be the external magnetic field, with a corresponding (nonlinear) magnetic susceptibility.

\subsection{Probe construction}

\begin{figure}
\includegraphics[height=8.5cm]{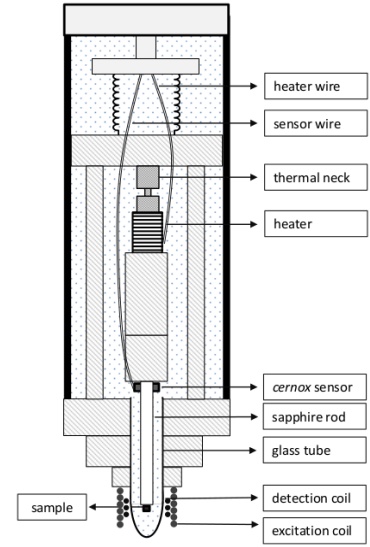}
\caption{Schematic depiction of the probe. High vacuum is maintained inside, and all junctions are either soldered or sealed with indium. The sample is placed on a sapphire rod in close vicinity to the cernox sensor, and heated through a copper element by a resistance wire heater. Thermal contact between the sample and environment is limited to the thin neck above the heater, and the lead wires are thermally insulated and immobilized.}
\label{shema}
\end{figure}

When measuring conductivity, the quantity of interest is often not its absolute value, but its temperature dependence. It was necessary for our apparatus to enable large variation in sample temperature, starting as low as possible. This was achieved by immersing the probe into a cryogenic liquid, which kept it at a constant temperature. The sample was placed inside the evacuated probe, and weakly thermally connected to the exterior. A heater situated in close proximity to the sample modified its temperature, while that of the excitation and detection coils, as well as lead conductors, was kept constant. We could therefore be sure that all thermal dependences are due to sample properties, and not an artefact arising from changing experimental conditions.

The probe consists of a closed thin-walled stainless steel tube inside of which high vacuum ($\sim$ $10^{-6}$ mbar) is maintained. The tip of the probe is sealed off with a test-tube made of electrically inactive quartz glass, as shown in Figure \ref{shema}. A sapphire sample holder is placed inside the tube, near the tip of the probe, so that it terminates inside the glass tube. The sample is attached to the sapphire rod by a thin layer of Apiezon N thermal grease, and the rod brought into contact with a copper element around which the heater is wound. A $cernox$ sensor is placed in the vicinity of the sample holder, and the thermal contact improved by thermal grease. 

A thin cylindrical piece of metal determines the thermal contact between sample holder and cryogenic environment. Depending on the temperature range of the experiment, this element can be exchanged. At higher temperatures ($\sim$50~K) we wish to slow down the cooling in order to preserve cryogenic liquid and speed up the measurement. Poorer thermal contact is thus preferable, and a thin stainless steel neck is used. In order to reach lower temperatures ($<$20~K) better thermal contact is achieved using a thicker brass link. We can lower the temperature of the sample even more by injecting exchange gas (helium) into the probe, and subsequently pumping it away.

The heater is controlled by a LakeShore Temperature controller, and the wires that connect the heater and sensor to the controller are thermally insulated and immobilized in the probe by PTFE disks. All junctions immersed in cryogenic liquid are sealed by indium wire. 

The excitation and detection coils are wound around the glass tube. Because of impedance mismatch effects in the RF range, the actual physical construction of the probe is somewhat more complex than the system proposed in Section 2. The excitation coil is wound from copper wire and immobilized by epoxy resin. A thin layer of silver paint is applied to one side and grounded by a copper wire, thus distributing capacity along the coil. \cite {zavojnica} The coil is terminated, and the effective scheme of the excitation circuit is given by a series of low pass filters, whose impedance is matched to that of the lead conductors up to 20~MHz. 

Signal detection can be either resonant or non-resonant, depending on the requirements and circumstances of the experiment. The resonant method is used when measuring low signals. Inside the excitation coil, and coaxial with it, a smaller coil of just a few turns is wound from manganin resistive wire. The coil is incorporated into a parallel $LC$ resonant circuit. Through a suitable choice of the matching capacitor, impedance is matched to 50 $\Omega$ at the resonant frequency, enabling any induced signal with that frequency to propagate into the lead conductors. The excitation frequency is then set to one-third or one-fifth of the resonant frequency of the $LC$ detection circuit, depending which harmonic is measured.

In the nonresonant method, the excitation circuit is simultaneously used for detection. Instead of terminating the coil, a coaxial cable is connected to it, and it works in transmission mode. Since the impedance is matched to 50 $\Omega$ in the range from DC to $\sim 100$~MHz, induced signals up to that frequency can propagate into the lead conductor with virtually no reflection (SWR$<$1.3). The nonresonant method is in general less sensitive than resonant, resulting in a lower signal-to-noise ratio. On the other hand, for sufficiently large signals this method is convenient, because the measurement procedure is exceedingly simple. It allows the change of frequency with no physical intervention on the probe, and also lacks any components sensitive to abrupt temperature change, such as surface mounted capacitors. Furthermore, it is preferable for measuring large or conductive samples, since a significant change of their properties might change both the natural frequency and quality factor of the resonant circuit.

The $LC$ circuit has been tuned to a desired frequency at room temperature. Since the resistivity of magnanin wire changes negligibly, both the frequency and Q-factor of the circuit remain relatively constant when immersed in the cryogenic liquid. The circuits' characteristics at high and low temperatures were analysed with a $Rohde\&Schwarz$ network analyser, and variations up to a few percent between room temperature and liquid helium (4.2~K) have been established. This implies that modifications to the probe would allow measurements in a variable temperature insert (VTI) which would reach significantly lower temperatures.

\subsection{Low power measurement procedure}
\label{low}

The conductivity of a material, especially its nonlinear component, is often dependent on signal strength. Valuable information on the underlying physical process can be gained from analysing this relation, which is why both fine and coarse excitation strength variation is illuminative. We developed two techniques that use the probe described in Section 3, and operate on significantly different excitation power scales.

\begin{figure}
\includegraphics[height=7cm]{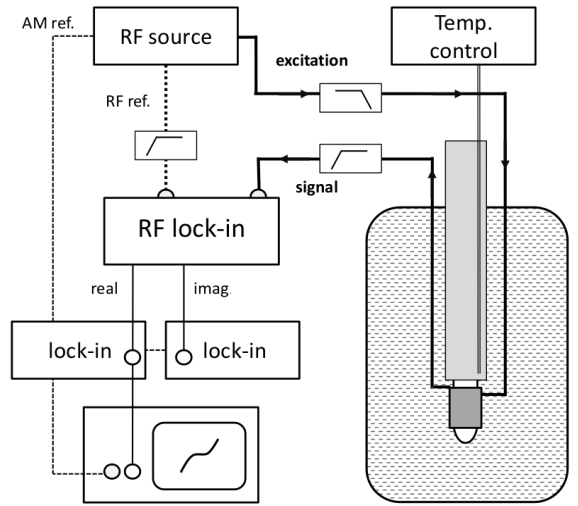}
\caption{Low power measurement procedure. The radio-frequency amplitude-modulated signal generated in the RF source is filtered and passes through the excitation coil before dissipating in the resistors that terminate the coil. Currents in the sample induce a signal in the detection circuit. The signal is filtered, demodulated and amplified in the RF lock-in amplifier. Furthermore, any background RF signal is eliminated by directing the two RF lock-in outputs into a pair of audio frequent lock-in amplifiers locked to the AM frequency of the signal. Thick lines represent radio-frequency, and thin lines audio-frequency signal. Dotted lines are signal reference paths.}
\label{shema-low}
\end{figure}

The low power measurement procedure is schematically depicted in Figure \ref{shema-low}. There is no difference in signal processing for the resonant and nonresonant detection method.

The excitation signal of frequency $\omega$ is generated by the radio-frequency source SRS DS345. The frequency is stabilized by an external 10~MHz reference provided by an HP 5342A frequency counter. An audio-frequency modulation is imposed on the amplitude in the whole range. In order to eliminate any higher harmonics from the source, the signal passes through a low pass filter before reaching the excitation circuit.

Signal induced in the sample is picked up by the detection circuit, whether resonant or non-resonant. The signal is led out of the cryostat by a separate semi rigid coaxial cable, and passes through a high-pass filter which eliminates the first harmonic. The elimination of the first harmonic before any amplification of the signal is essential, since any distortion of it would pollute the higher harmonic signals. An SRS844 radio-frequency lock-in amplifier acts as a detector and demodulator. It is locked to the TTL output of the source, passing through a custom frequency multiplier. The scheme of the frequency multiplier is similar to the one proposed by Wenzel,\cite{multiplier} and the values of the components needed for the required frequency range determined by simulation in the free software TINA.\cite{Tina} A high pass filter could be used instead of the multiplier, but in our experience this leads to higher phase noise and lower locking quality.

The two outputs of the lock-in amplifier are proportional to a combination of the real (in-phase) and imaginary (quadrature) component of the incoming signal at the reference frequency. However, to further increase the signal-to-noise ratio (SNR), the signal undergoes one more stage of amplification. Separately, the signals from the outputs are brought to two SRS830 lock-in amplifiers, whose reference is the amplitude modulation frequency given by the source. Using standard GPIB communication, this signal, along with the temperature data provided from the LakeShore 336 Temperature Controller, is analysed and plotted in real time using a custom made program written in $LabView$. 

An oscilloscope operating in the $X-Y$ mode is used as a visual control tool in the course of the measurement. The instantaneous amplitude of the driving current is displayed against the value of the RF signal, allowing the detection of any change or irregularity of the probe.

The RF lock-in amplifier allows four different output options: the real and imaginary component, absolute value and phase value. Nominally, it is irrelevant which output channels are measured, since the information contained in any pair of them is equivalent. Actually, the phase value tends to fluctuate more when measuring small signals, meaning that only the absolute value can be analysed reliably. In most experiments, our choice of outputs were the absolute value and the imaginary component, which was set to zero at high temperatures, by adjusting the RF reference phase. If the imaginary component is not negligible, the remaining offset of the absolute value cannot be simply subtracted. The relation which connects the measured value $R$, which includes the offset $R_0$, the imaginary component $y$ and the absolute value of the signal $R_{sig}$ is:
\begin{equation}
R_{sig}=\sqrt{\left(R_0-\sqrt{R^2-y^2}\right)^2+y^2}
\end{equation}
We emphasize that this formula is valid only if the offset of the imaginary signal component $y$ is zero, otherwise it has to be taken into account as well. 
 
\subsection{High power measurement procedure}
\label{high}
\begin{figure}
\centering
\includegraphics[height=7cm]{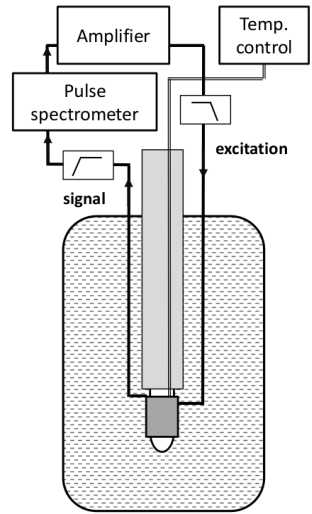}
\caption{The high power measurement procedure. A microsecond pulse is generated in the spectrometer, amplified and filtered afterwards, so that only the first harmonic passes through the excitation circuit. The signal induced in the detection circuit is again filtered, and analysed in the spectrometer simultaneously with the pulse.}
\label{shema-high}
\end{figure}

Because of sensitive electronic components and limited RF source power, the technique described above allows excitation strength variation up to $\sim$0.5 W. An NMR pulse spectrometer was used for stronger signal excitation and detection, as shown in Figure \ref{shema-high}. 

The spectrometer generates a pulse which is amplified up to the power of $\sim$150 W by a CPC 5T1000M MRI amplifier. It was necessary to limit the signal to short pulses in order to prevent damage to the probe. Microsecond ($\sim$20 $\mu$s) pulses were sent every 8 ms, setting the RMS power delivered to the excitation circuit to cc. 3 W. Due to the nature of NMR measurements for which the amplifier is designed, the spectral distribution of the output is not pure enough for our purpose. Therefore, a custom made high power 7-pole Chebyshev low pass filter was constructed to eliminate higher harmonics before the signal arrives to the probe. The rejection of the $3\omega$ signal is typically around 80 dB.

The signal induced in the probe is filtered, integrated and analysed in the spectrometer simultaneously with the pulse. Typically $\sim$ 10000 pulses were acquired and averaged for every measurement point. The long duration of each measurement makes the slowly changing offset originating from the amplifier significant. In order to eliminate it, data acquisition was performed in a wide span of strengths, ranging from 10$\%$ to 90$\%$ of full pulse strength. The signals measured at different temperatures were normalized to a few lowest strength values, where the signal was presumably small compared to the offset, thus compensating offset fluctuations. 

The high-power technique is more sensitive than the low-power one by a factor of 100 to 1000, but the measuring procedure is long lasting, complex and requires a considerable amount of data processing, in contrast to the straightforward low-power experiment.

\section{Measurement and validation}
\label{measurement}
\subsection{Low-power measurement}
\begin{figure}
\includegraphics[height=6.5cm]{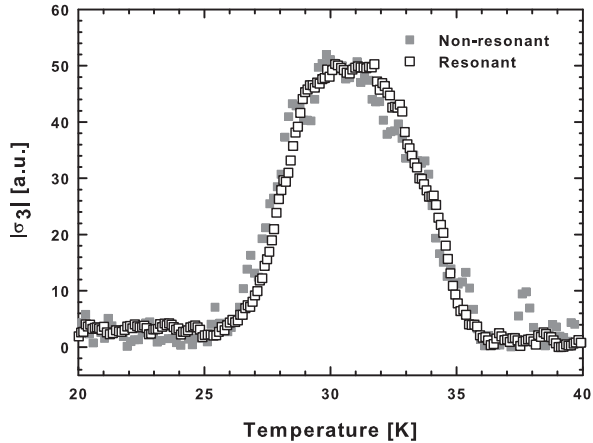}
\caption{Normalized low power third harmonic measurements of LSCO. The agreement between non-resonant (full squares) and resonant (empty squares) method is very good, with the noise being appreciably lower in the resonant method.}
\label{LSCO}
\end{figure}

In order to test our apparatus, measurements were performed on a millimeter-size single crystal of La$_{2-x}$Sr$_x$CuO$_2$ (LSCO) in a wide range of temperatures. It was chosen because nonlinear response was previously detected in high-Tc superconductors at or around the superconducting transition temperature at various frequencies, in both crystalline and ceramic samples. Depending on sample type and excitation frequency, nonlinear signals are attributed to various effects, such as intergranular Josephson coupling,\cite{Jeffries1988} fluxon dynamics,\cite{Leviev1991} BKT transitions\cite{Kuzmichevphysica, Kuzmichevtt} and order parameter perturbation.\cite{Leviev1989, Agliolo1997} A significant problem, extensively discussed in literature, is the onset of nonlinear signals on account of fluctuations of sample temperature.\cite{Mishonov2002, Cheenne2003} Any current passing through the sample inevitably causes heating, but if the current frequency is sufficiently high the sample temperature will not follow its oscillations - the heating will be uniform in time. The characteristic frequency where the current variation is too fast for the sample temperature to follow it is determined by the physical properties of the sample: its characteristic dimension $L$, density $\rho$, heat capacity $c_p$, and thermal conductivity $\kappa$. An estimate derived by dimensional analysis from the heat equation suggests this frequency to be
\begin{equation}
\omega_T \sim \frac{\kappa}{\rho c L^2}.
\
\end{equation}
If the excitation frequency is significantly larger, nonlinear heating effects will be negligible. The value of $\omega_T$ for our sample is $\sim$kHz, which implies that radio-frequency currents cause negligible thermal fluctuations. Throughout the experiment, for typical temperature ramp rates of 3-5~K/min, the absolute precision of the sample temperature was $\sim$~1~K, but greater precision could be achieved by lowering the ramp rate. 

Results obtained by resonant and nonresonant detection are compared in Figure \ref{LSCO}. The third harmonic signal peaks around $T_c$, with a better SNR for the resonant method, as expected due to its higher Q factor. The fact that both signal peak and shape coincide regardless of detection method indicates that the nonlinear response arises as a result of intrinsic physical processes in the sample, and not as an artefact of the technique. Specifically, it shows that the properties of the resonant circuit do not change significantly when the sample undergoes a superconducting transition. Systematic analysis of the higher harmonic signals induced in various high-Tc superconductors will be published elsewhere.

\subsection{High-power measurement}
\begin{figure}
\includegraphics[height=6.5cm]{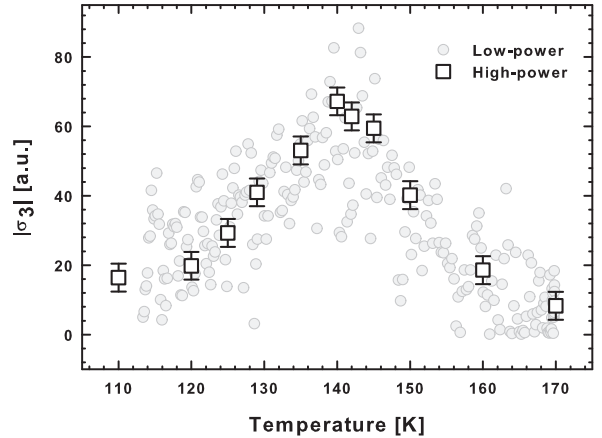}
\caption{Normalized third harmonic measurements on an undoped iron pnictide. Open squares denote results obtained by the high-power method, and a sharp peak is visible at $T_n$=138~K, which matches the nematic phase transition temperature. The low-power measurement results, shown here as circles, also peak at $T_n$, but have a significantly lower SNR. }
\label{pniktid}
\end{figure}

Except for the superconducting transition, it is possible to detect other, more subtle, electronic phases present in strongly correlated materials by measuring nonlinear effects. In order to investigate the sensitivity of our apparatus, we measured the third harmonic response on an undoped ($x=$0) sample of Ba(Fe$_{1-x}$Co$_x$)$_2$As$_2$, an iron pnictide, which is known to undergo a nematic phase transition at 138 K.\cite{Chu2012} Measured by the low power method, the  third harmonic response shows a feature around the transition temperature, but no details can be discerned, as the SNR is low. High-power measurements where therefore conducted, and the results are shown in Figure \ref{pniktid}. A sharp peak is confirmed at $T_n\sim$140~K.

 Contrary to the low power method, which displays the third harmonic response values in real time, the pulse method requires data processing. Furthermore, in addition to acquisition time of around 1$h$, it was necessary to wait for the temperature to stabilize before each measurement, so the total time needed for each high-power data point amounted to $\sim$ 90 minutes. Therefore, despite obvious advantages of the high-power method regarding the SNR, it is not always preferable to the low-power method.
 
 Ultimately, the choice of the measurement method must be suited to the nature of the experiment and signal strength. The ``what you see is what you get'' approach makes the low power method ideal for wide temperature range scanning and quick measurements of sufficiently large signals, as well as coarse measurements of the third harmonic  dependence on a variable parameter, such as an external magnetic field. The high power method is suitable for measuring small signals, as well as physical processes which only occur if the excitation exceeds a threshold value (such as pinning effects).

\section{Conclusion}

We developed a probe that enables simple and precise measurements of the nonlinear conductivity of small samples in a wide range of temperatures. Two methods of detection have been put forward: one resonant, which is more sensitive, and the other non-resonant, suitable for larger signals and samples. Furthermore, two distinct methods of excitation generation and data acquisition have been proposed, suitable for different power scales. The low power measurements use the continuous signal from an RF source as excitation, and lock-in amplifiers as detectors. The high power method uses a commercial NMR spectrometer and accompanying amplifiers. Compared to conventional DC conductivity measurements, the method presented is nondestructive, simple to operate, and in case of low power measurements, faster. The major problems of existing methods -- heating effects for low-frequency measurements, and skin depth issues leading to sensitivity to surface effects in microwave experiments -- are avoided in our design. A specific, important application is the nonlinear response of high-Tc materials, which was successfully demonstrated on a LSCO single crystal close to its superconducting $T_c$, and an iron pnictide around its nematic transition. Since the high-Tc systems display a variety of electronic ordered phases, our probe enables valuable insight into the structure of the phase diagram of such materials.  Although the apparatus was designed with single crystal high-Tc superconductors in mind, it can be applied to a wide scope of interesting physical systems, including CDW, novel magnetic materials, spin-frustrated materials, and the like.

\section*{Acknowledgements}

We would like to thank N. Barišić for providing the LSCO single crystal, and IFW Dresden for the iron-arsenide sample. Also, we gratefully acknowledge the assistance of A. Dulčić and I. Živković. This work was supported by the Croatian Science Foundation (HRZZ).

\bibliography{nonlinear_conductivity}
\end{document}